# Actinide and Lanthanide Dioxide Lattice Dilatation Mechanisms with Defect Ingrowth

Seçkin D. Günay

*Department of Physics, Faculty of Science, Yıldız Technical University, Esenler, 34210, İstanbul, Turkey*

Corresponding author. Tel.: +90 212 3834289 E-mail address: sgunay@yildiz.edu.tr

# Actinide and Lanthanide Dioxide Lattice Dilatation Mechanisms with Defect Ingrowth

**Abstract**

α-particle irradiated single-crystal isostructural oxide lattices ($CmO_2$, $AmO_2$, $CeO_2$, $UO_2$, $NpO_2$, $PuO_2$, $ThO_2$) were modeled by molecular dynamics simulation method. Lattice parameter changes with IFP cation defects displayed exponential increases. These data were compatible with the available lattice parameter changes with α-particle dose experiments. Pre- and post- peaks emerged around principal peaks of irradiated oxide cation-cation radial distribution functions, which indicate obstruction and distortion type defects, respectively. Dependence of lattice dilatations on the number of obstruction type cation defects was examined. In a previous study, obstruction type uranium defects were found to be directly responsible for the $UO_2$ lattice swelling and there was a linear relationship between them. It was also determined that this linear equation has two different slopes at low and high defect concentrations. In this paper, it was found that these phenomena were not specific to $UO_2$ and applicable to all fluorite-structured actinide and lanthanide dioxides studied here. These findings provide clues about the existence of more general law.

## 1. Introduction

Studies on irradiated materials with alpha particles have been done in the literature for decades. To the best of our knowledge, a significant portion of these studies has been done on isostructural oxides ($CeO_2$, $UO_2$, $PuO_2$) [1-4]. In this study, lattice dilatation mechanisms and defect ingrowth of some single-crystal fluorite-type isostructural oxides (actinide and lanthanide dioxides, i.e., $CmO_2$, $AmO_2$, $CeO_2$, $UO_2$, $NpO_2$, $PuO_2$, $ThO_2$) due to α-particle irradiation were examined by classical molecular dynamics simulation method and compared with available experimental results. The effect of α-particle irradiation on the crystal was modeled by the techniques used in previous studies [5,6]. According to these, the obstruction type (OT) defect was directly responsible for the change in the $UO_2$ lattice parameter. There was a linear relationship between the number of OT uranium defects and the fractional change in uranium dioxide lattice parameter. Moreover, an increase in the slope of the linear equation was observed above a critical number of OT uranium defects. The existence of a similar relationship will be questioned for other isostructural dioxides ($CmO_2$, $AmO_2$, $CeO_2$, $NpO_2$, $PuO_2$, $ThO_2$) and how these defects change the physical properties of molecules will be investigated by considering previous outcomes. The dependence of obtained results [5] for $UO_2$ on input parameters was examined by changing the interaction potential and increasing the number of atoms in the simulation box. In addition to these studies, the amount of lattice dilatation caused by self-irradiation of some actinide dioxide materials ($CmO_2$, $AmO_2$, $PuO_2$) [7-13] as a result of α-decay was compared with the obtained data. Although no comparisons had been made with other self-irradiated actinide dioxides (($U, Th)O_2$, MOX) [14,15], discussions were also performed on these molecules.

## 2. Methods

In this study, a pair potential with the many-body embedded atom model was used to model the interactions between ions.

$$E_i = \frac{1}{2}\sum_j \phi_{\alpha\beta}(r_{ij}) - G_\alpha \sqrt{\sum_j \sigma_\beta(r_{ij})}$$

The first part is the pair-wise interaction term and the second term is the many-body term. The pair-wise term consists of parts such as long-range Coulomb, Morse and Buckingham interactions. The cut-off distance was determined as 11.0 Å for short-range interactions potentials. Potential parameters for some isostructural oxides ($CmO_2$, $AmO_2$, $CeO_2$, $UO_2$, $NpO_2$, $PuO_2$, $ThO_2$) were tabulated by Cooper et al.[16].

Molecular dynamics simulation supercell boxes (10×10×10) were constructed for these molecules. Alpha-irradiation damage was modeled by randomly moving cations in the crystal structure. Cations were relocated from their normal position to interstitial sites which are called initial Frenkel pairs (IFP). The simulation was initialized when a new cation IFP defect was introduced in the simulation box. Thus, the number of defects was increased one by one to reach the maximum value of 140 IFP. This value was chosen to observe the saturation point that crystal lattice could expand as a result of the radiation.

In experiments conducted with α-particle radiation [1-3], particles acting on one side of the structure equally affected the entire crystal. In order to model homogeneous irradiation from one side of the crystal, all cations were shifted in the Y-Z layer and were not placed in the successive layers. At some point, for a high number of IFPs, the artificially created defects were so close to each other that it was not possible to comply with this method.

Since this artificial positioning of cations adds a large amount of energy in the "irradiated" crystal structure, the simulation was initiated by an energy minimization technique. Steepest descent algorithm was used for energy minimization. The molecular dynamics simulation was initiated after this process. Supercells were annealed for 100ps at 300K temperature. Isobaric-isothermal (NPT) ensemble was applied to the simulation box. The constant stress method, which was implemented on the MD simulation box, allowing the change in size and shape of the supercell. The system was exposed to a heat bath, by so-called Nose-Hoover thermostat, to keep the temperature constant. Lattice constants, atomic coordinates, radial distribution functions, etc. properties were saved to calculate physical properties like the number of defects and swelling.

## 3. Results

Fission and α-particle irradiation effects and thermal recovery of uranium dioxide were investigated by the present author in previous studies [5,6]. In this study, calculations were further expanded based on those findings and the effect of α-particles on isostructural oxides. Numerous experiments have been conducted on fluorite-structured compounds, which were exposed to α-particle radiation or α-decay events. Damage ingrowth, swelling, formation of defects, lattice recovery and damage annealing were mainly investigated in some actinide dioxides, i.e., $CmO_2$ [8], $AmO_2$ [9, 10], $PuO_2$ [11-13], $UO_2$[1-3], $(U, Th)O_2$ [14], MOX [15]. Except for some experiments in which Weber [1,2] used single-crystal $UO_2$ materials, other experimental groups utilized polycrystalline actinide dioxide materials.

### *3.1 Lattice Parameter*

#### *3.1.1 Changes in the Lattice parameters*

Artificial defects called IFP (Initial Frenkel Pairs) prepared by random displacement of cations in the simulation box. Each point in Fig. 1 corresponds to an independent simulation of 100 ps initiated by so-called irradiated and energy minimized supercell boxes. The variation of fractional changes in lattice parameters ($\Delta a/a_0$) with the initial number of Frenkel pairs (IFP) is shown in Fig. 1. Exponential growth was observed for all isostructural oxides. Our calculations resemble the experimental results, which was the fractional change in lattice parameters as a function of α-particle radiation doses of single- and poly-crystal fluorite-structured actinide dioxides [1,3]. The saturation points for oxides from large to small values can be listed as follows: $CmO_2$, $AmO_2$, $CeO_2$, $UO_2$, $NpO_2$, $PuO_2$, $ThO_2$.

#### *3.1.2 Saturation Point*

Experiments show that the saturation level of fractional change in lattice parameter varies depending on the type of radiation. The saturation levels of single-crystal $UO_2$ in descending order are α-particles (~0.8%), α-recoil (~0.4-0.5%) and fission (~0.1%) [1,14]. The main reason for this difference between saturation points is the number of defects emerged for each type of radiation. Primary atomic displacements per interacting particles were listed as, 100 for α-particle irradiation, 1500 for α-decay and 100000 for fission events [14]. Moreover, depending on the type of defects, the life span in which the defects are created and annihilated would have a significant impact on the saturation level [2,14]. As it is understood from the previous discussions [14], it was commented that the α-particles have lower removal rates of defects from the lattice by recombination, annihilation or clustering than other radiation types. In some studies, it had been mentioned that thermal spikes of fission and α-recoil events may contribute to the recombination or cluster formation of defects [1,4]. Finally, it was commented that defect annealing and clustering may take place along the track of α-recoil atoms.

The data obtained from our previous studies [5,6] provide information about the defects responsible for the irradiation-induced swelling in uranium dioxide crystal lattice structures. It was concluded that the amount of lattice dilatation depends on the number of obstruction type (OT) defects. These defects are uranium and oxygen ions for α-particle radiation and fission events, respectively. The saturation level of lattice expansion for α-recoil-damaged uranium dioxide structure is between α-particle and fission damage saturation levels [1]. It could be concluded that thermal spikes play a key role in the difference among the saturation levels of these three events and lead to a significant change in the amount of cation obstruction type defects. In the fission event, based on the previous findings [5], it could be estimated uranium atoms are displaced from their normal positions to form OT uranium defects but almost immediately these defects annihilate or recombine as a result of thermal spikes. Thus, a limited amount of swelling occurs with the OT oxygen defects remaining in the lattice [5]. The saturation level in the α-recoil event demonstrates that the OT uranium defects are partially recombined as a result of thermal spikes. The thermal effect is not sufficient to eliminate all defects or the effect is limited to certain regions. The recombination of defects as a result of thermal spikes and the formation of defects as a result of radiation occurred simultaneously, and an intermediate saturation level for α-recoil event took place higher than the fission event and lower than α-particle radiation. Thermal spikes also occur in the case of α-particle radiation, but since this occurs in a limited number, it eliminates a small amount of OT defect and slightly reduces the increase in lattice constant.

The saturation points of single and polycrystalline structures calculated by experimental and classical MD simulation methods are listed in Table 1. Simulation results of single-crystal structures of oxides in the second column were acquired from Fig. 1 and compared with Weber's experiments [3] on α-particle irradiated polycrystalline materials in the third column. $\Delta a$ and $a_0$ are the change in the unit cell size and the lattice constant of the perfect crystal, respectively. Although the second column is the single-crystal and the third column is the polycrystalline structure, results show that there is a correlation between both data groups and decreases from top to bottom. A similar study [5] previously conducted on the 8×8×8 single-crystal $UO_2$ supercell with another interaction potential [17,18] yielded ~1.05%, which is very close to the results attained in the current study (~1.03%).

The amount of swelling in the single-crystal material is higher than the polycrystalline material. Experimental saturation point data are 0.84% and 0.58% for single-crystal and polycrystalline $UO_2$, respectively. Single-crystal data is about 45% higher than polycrystalline data. Since there are no experiments of α-particle irradiation effects on the single-crystal structures of other oxides in the literature, one can make simple assumptions about what results would be achieved if α-particle irradiation experiments were performed on these single-crystal structures, assuming similar relationships between single and polycrystalline structures as in $UO_2$. Saturation points of α-particle irradiated single-crystals could be found roughly by using saturation point values of α-irradiated polycrystalline isostructural oxides ($CeO_2$, $PuO_2$) in Table 1.

There are other experimental data (Table 2) on actinide oxides ($CmO_2$, $AmO_2$, $PuO_2$) in the literature. These are self-irradiation from α-decay, which exhibits both α-particle irradiation and the α-recoil event. There is a distribution in the experimental saturation point data, and they are not consistent among themselves. For example, as shown in Table 2, saturation points of the lattice parameter changes for polycrystalline $PuO_2$ [15] are between $2.83 \times 10^{-3} - 3.90 \times 10^{-3}$. Therefore, a similar relationship as in Table 1 does not exist in Table 2 and no comparison can be made. The main reason for this is thermal spikes caused by α-recoil atoms and consequently the random annihilation of cation defects. Data from our previous studies [5, 6] demonstrated the importance of the effect of cation Frenkel (OT) defect pairs. The uncertainty in the number of thermal spikes and the amount of temperature increase in the crystal makes it difficult to predict the number of annihilated OT cation defect pairs and the corresponding swelling. Besides all these, the dilation in lattice constants is sensitive to the preparation conditions of samples [14]. Therefore, there is no significant relationship between our and experimenters' data, similar to Table 1.

It would also be interesting to compare the experimental and the MD simulation results for α-particle irradiated single-crystal $UO_2$ in Table 1. Single-crystal data from the experiment (0.84%) is somewhat different when compared to single-crystal MD simulation data (~1.03%). In order to clarify this situation and explain the difference between results, a simple computer experiment can be made as follows: Cations were artificially placed and energy minimization was initiated in the calculations made with a standard MD simulation so far. However in a real experiment, as a result of α-particle radiation, atoms are dislocated from their normal positions and moved into an unstable state. This adds a large amount of kinetic energy to the simulation box. Analogously, to observe a non-equilibrated (unstable) simulation box with excess energy, the simulations were initiated without energy minimization. The obtained data from MD gives us a chance to roughly compare it with experimental data. Simulation box systems with and without energy minimization were compared and significant differences were observed between saturation points (Fig. 2). Saturation point data for the MD simulation supercell box without energy minimization

(0.74%) were obtained close to experimental data (0.84%). Comparing these two results may be speculative. However, it is an important example in terms of seeing the effect of excess energy (e.g. thermal spikes in an experiment) on lattice constant and defects.

*3.1.3 Effects of cation defects on lattice parameters*

The formation of cation defects as a result of α-particle irradiation has a primary effect on the dilation of the lattice parameter. As observed in the previous study [5], there is a linear relationship between the number of interstitial (OT) uranium defects and lattice swelling. This interesting result remained valid even if the number of atoms in the simulation had been changed [5].

Once again, the number of atoms in the simulation box was increased (10×10×10) and a new interaction potential was used in this study. The sensitivity of the previously obtained data to simulation inputs (atom number, interaction potential) was tested for uranium dioxide. The presence of a similar relationship in other oxides was also investigated here.

In spite of the changes in input parameters, linearity still exists between the number of obstruction type uranium defects and the fractional change in the $UO_2$ lattice parameter in Fig. 3. The fractional change in the lattice parameter ascended with the increasing number of OT uranium defects and the linear relationship was maintained up to 0.48%. The gradient of the graph changed above this point and a new linear equation was achieved with a steeper slope value. The new equation ended up around a maximum value of 1.05%. Simulations were also carried out using other oxides: $CmO_2$, $AmO_2$, $PuO_2$, $CeO_2$, $NpO_2$, $ThO_2$. A two-part linear equation was also observed for each of these molecules (see Fig. S1-S6 in the supplementary material). Linear equations were fitted separately to each (first and second) part of the data (Fig. 4) and slopes were tabulated in Table 3. Thus, this interesting result, which applies to uranium dioxide, have also been achieved in other fluorite-type actinide and lanthanide dioxides.

Table 3 presents the primary and secondary slopes for all the oxides investigated in this study (Fig. 4). The difference between the slopes of first and second parts of the function in $ThO_2$ is very small. Therefore, the dependence of the number of OT cation defects on the fractional change in $ThO_2$ lattice parameter can also be expressed with a single linear equation.

Table 3 also presents $\Delta v_F$ volume increments per OT cation defects which was attained from primary and secondary slopes. The equation for volume increment was given as follows,

$$\frac{\Delta a}{a_0} = \left(\frac{\Delta v_F}{3 n_{cell} V_0}\right) N_F \qquad (1)$$

$(\Delta a/a_o)/N_F$ is the slope of the fractional change of lattice parameter versus OT cation defects (Fig. 4). $V_o$ is the unit cell volume of the undamaged crystal and $n_{cell}$ is the number of unit cells in the MD supercell. Although there were many defects in the irradiated lattice, only OT cation defects were taken into account in determining volumetric increases. Thus, unlike previous studies [1,3] only one type of defect is taken into account but not all defects while using the same equation. Therefore, it will not make sense to compare the results of this study with the experimental results due to differences in the definition of expression (supplementary material).

In the previous study [5], 46.62 $Å^3$ and 64.97 $Å^3$ were acquired from the primary and secondary slopes of $UO_2$ using the same calculation method. In the present study, the number of atoms was increased and a different interaction potential [16] was implemented into another program. Former results are compatible with this work for $UO_2$. There is a difference between $\Delta v_F$ data calculated from the primary and secondary parts of the graph for all oxides. The difference in $ThO_2$ is almost negligible (4.9 $Å^3$). It is small for $NpO_2$ but too large to be neglected (9 $Å^3$). Calculated data are larger than others and close to each other (14-20 $Å^3$) for $CmO_2$, $AmO_2$, $PuO_2$, $CeO_2$, $UO_2$ (Table 3).

Breakpoints and maximum points were acquired from lattice parameter changes with OT cation defect number data for each of the oxides (see Fig. S1-S6 in the supplementary material and Fig. 3). PuO$_2$ molecule has the highest number of OT cation defects (n=54) without any change in the first part of the linear function among other oxide molecules (Table 4). Besides, when the lattice swelling reaches its maximum point, it is the molecule with the most OT cation defect possessing lattice (n=101) compared to other molecules. However, the maximum swelling effect of OT plutonium defects on its lattice is moderate (%1.01) compared to the maximum effects of the OT cation defects of other molecules on their lattices. Note that the maximum point differs from the saturation point due to dispersion in the data.

In order to compare the defect-bearing capacities of oxides, it is necessary to calculate the OT defect number density. This value is acquired by dividing the OT cation defect number to the volume of the dilated lattice. When the OT cation defect-bearing capacities of oxides were compared for the saturation level, PuO$_2$ molecule had a larger value than others, and oxides were listed from large to small values, respectively, as follows: PuO$_2$, CmO$_2$, AmO$_2$, NpO$_2$, UO$_2$, CeO$_2$, ThO$_2$ (Fig. 5). In Weber's study [3], the number of defects per unit cell for polycrystalline structures was calculated (PuO$_2$, UO$_2$, CeO$_2$) and it is consistent with the order in this calculation. The types of defects mentioned in both studies are not the same, but it is important to mention when discussing the defect-bearing capacities because the increase in the number of OT defects in the lattice leads to an enhancement in the number of other defects [5,6].

### 3.2 Radial Distribution Functions and Visual Observations

Fig. 6 and 7 represent cation-cation radial distribution functions (RDF). The RDF designates how the density changes as a function of the distance around a central reference atom. Significant peaks with precise boundaries indicate that atoms are stationary in the crystal and retain their solid structure. The effects of atom configurations on RDF data were examined by considering two different groups, 30 IFP and 100 IFP. These cation defects were manually changed and the RDFs were obtained at the end of 100ps for isostructural oxides (Fig. 6 and 7). These two groups represent samples before (primary slope) and after (secondary slope) the breaking point in the linear equation (Fig. 4). Both samples investigated here also correspond to low and high doses of alpha irradiation. All of these aspects were taken into account when examining the effects of defects on RDF in the following subsections. The OT (obstruction type) defect, DT (distortion type) defect, pre-peak, post-peak, principal peak, secondary peak are important terms when referring to atoms and RDF data. Pre- and post-peaks that do not exist in a perfect crystal structure around principal peaks were observed in irradiated isostructural oxide RDF graphs. Both 30 IFP and 100 IFP samples were discussed in each of the pre- and post-peak subsections. These peaks were examined from two different perspectives; peak position of an oxide relative to peak positions of other oxides and peak position of an oxide relative to its principal peak position.

*3.2.1 Pre-peaks*

Pre-peaks, which are not normally found in undamaged crystal structures, have emerged markedly in the range 2.5Å-3.5Å for all oxides in this study (see Fig. 6 and 7). This proves us that there is a new interstitial cation defect position from a reference cation atom for all isostructural oxides in this study which is closer than the nearest neighbor cation-cation distances (principal peak in Fig. 6 and 7). The highest point projection of the pre-peak on the x-axis is the most likely location where the interstitial defect is located. The pre-peak heights $(g(r) \approx 0.2 - 0.4)$ are small compared to both principal peak levels $(g(r) \approx 11 - 15)$ and the average cation density level $(g(r) = 1)$ in the RDF graph (Fig. 6 and 7).

A similar peak was observed in uranium dioxide [5], and it was proved that interstitial (OT) uranium defect was trapped in an octahedral cage which was coordinated with six other uranium ions. These defects can easily be seen by visual observations of crystal structures from the <110> direction. The instantaneous OT defect position and closest cation distances for the $UO_2$ molecule, which corresponds to 30 IFP sample U-U pre-peak (Fig. 6), found in this study are given in Fig. 8a. Analogously, in all oxide molecules investigated in this study, OT cation defects were surrounded by six other cations. These six cations will also be referred to as DT defects in the following section. OT uranium defects - uranium distances over time were examined and the mean value was found to be consistent with the pre-peak in RDF data (see Fig. S7 in the supplementary material). When the distances were examined it is seen that they were slightly different from each other, that is, the octahedron has an asymmetric structure. In this particular example, the average value of three of the six distances is about 3.08Å, and the other three is about 3.01Å. This asymmetry occurs differently for each OT defect.

Although all oxides investigated in this study generally exhibit similar behavior in RDF, $ThO_2$ for 30 IFP (Fig. 6) and $PuO_2$, $ThO_2$ for 100IFP molecules (Fig. 7) differs slightly from others. All pre-peaks have clear boundaries in 30 IFP radial distribution function. Therefore, the number of defects can be calculated precisely without dispersion in data. The $ThO_2$ pre-peak is somewhat higher than others and shifted to the right (~3.12Å). The average position of the peaks of other molecules is around 3Å.

For 100 IFP, the $PuO_2$ pre-peak is broader and has less prominent boundaries than pre-peaks of other oxides. The lack of clarity of $PuO_2$ pre-peak boundaries increase uncertainty in calculations and make it challenging to determine the number of OT defects $\overline{M_\alpha}$ because it is crucial to set an upper and a lower limit on the x-axis for the equation [5]. As in 30 IFPs, the RDF pre-peaks of 100 IFP samples for $ThO_2$ are located on the right-hand side of other pre-peaks. The $ThO_2$ molecule has a steeper pre-peak with more evident boundaries. Thus, the number of defects for ThO$_2$ can be obtained more precisely with less uncertainty than other molecules.

*3.2.2 Post-peaks*

The emergence of post-peaks in Fig. 6 and 7 indicates DT cation defects. The main reason for the occurrence of these defects is OT cation defects. Aligned cations in their normal positions are disturbed as a result of the presence of OT cation defects. There is a cage consisting of six cations around OT defects, which forms an octahedron structure (Fig 8 and see Fig. S8-S13 in the supplementary material). These six cations in their normal positions are pushed diagonally (four cations on the horizontal axis) and perpendicularly (two cations on the vertical axis) into the <110> channel due to OT cation defects. Irregularities in the alignment of cations can be noticed by visual observations (Fig. 8a). Fig. 8b and 8c display DT defects that were driven out of their normal positions, and their distances relative to each other. Their effects can also be observed on RDFs with the emergence of new peaks (post-peaks) between 4Å-4.55Å (Fig. 6 and 7), which is consistent with twelve instantaneous interatomic distance values between DT cation - DT cation defects (Fig. 8b and 8c).

Despite the occurrence of pre- and post-peaks (Fig. 6), and change in all other parts of RDF data (inset in Fig. 6), 30 IFP sample data resemble the crystal structures. This would change with increasing α-irradiation doses and acquire a liquid-like appearance. Post-peaks with evident boundaries are seen in RDF graphs for each of the 30IFP oxide samples in the range of 4Å-4.55Å, which corresponds to low α-irradiation doses. Noticeable peaks in Fig. 6 indicate uniform distance distribution of distortion type defect locations. The post-peak of PuO$_2$ adjoins to its principal peak. Post-peaks of other oxides are relatively distant from their principal peaks when compared with PuO$_2$. Further, post-peaks of PuO$_2$ and ThO$_2$ are located to the right and left of the average position of other oxide molecules post-peaks.

Distortion type defects for 100 IFP samples (Fig. 7), which correspond to high α-irradiation doses, resemble atoms with thermal vibrations that are exposed to high temperatures due to their misalignment. RDF graphs show that the cation sub-lattice has a liquid-like appearance even at room temperature (inset in Fig. 7). The most important indicators of this phenomenon, when compared to an undamaged crystal, are broadening and decrease in heights of RDF peaks, and the fact that RDF data is not zero above 2.5Å.

Isostructural oxides with 100 IFP samples can be roughly divided into three groups by analyzing RDF data between 4Å-4.55Å. It is impossible to determine a precise location for DT defects in the $PuO_2$ molecule which belongs to the first group. The probability of being found from the edge of the principal peak (~4.2Å) towards 4.55Å is gradually decreasing. In the second group ($UO_2$, $CmO_2$, $AmO_2$, $CeO_2$, $NpO_2$), peaks appear vaguely in the 4.2Å-4.55Å range. In the third group, $ThO_2$ post-peak (4.3Å-4.55Å) is visible.

Another post-peak was also observed in the range 4.55Å-5.0Å. As the visual data is examined, it was determined that these peaks were related to OT cation defect - cation distances (Fig. 8d). Therefore, it is more appropriate to call these "pre-peak of secondary peak" (Fig. 6 and 7). OT defect - cation distance values in Fig. 8d verify RDF data. All these pre-peaks of secondary peaks are visible for 30 IFP samples, whereas only $ThO_2$ and $UO_2$ pre-peaks emerge in 100 IFP samples.

**4. Discussion**

In previous studies [5,6], the dependence of the change in α-particle irradiated $UO_2$ lattice parameter on the number of OT uranium defects was examined and it was shown that there was a linear relationship between them. It was also observed that this linear dependence consists of parts with two different inclinations. In this study, this dependency was tested by increasing the number of atoms in the MD simulation box, implementing a different interaction potential and employing a new program. Despite all these changes, the results obtained previously did not change. Another important question in this study was whether there was a similar incident in other fluorite-structured actinide and lanthanide dioxides ($CmO_2$, $AmO_2$, $CeO_2$, $NpO_2$, $PuO_2$, $ThO_2$). It was interesting to determine the presence of this phenomenon in other oxides. This means that a general statement could be written about the fluorite-structured actinide and lanthanide dioxides. A definition for all fluorite-structured dioxides or other fluorite-structured molecules can now be speculative and further research is needed. It should also be emphasized that slopes of first and second parts of the $ThO_2$ linear equation are almost equal to each other that the whole equation can be assumed to be one-part.

      The dependence of some single- and poly-crystal fluorite-type oxide lattice dilation on α-particle irradiation doses displayed exponential increases. Similarly, the expansion of single crystal oxide lattices, investigated in this study with the MD method, relative to cation IFP demonstrates exponential dependencies. Exponentially increasing functions reach saturation points and it is essential to compare calculated saturation points with experimental data. Although structures of interest in this study are single-crystal oxides, materials used in experiments are generally polycrystalline oxides except for one study with $UO_2$. Despite this difference, a correlation can be established between MD simulation and experimental data. When the saturation levels of single-crystal materials obtained from the simulation were sorted, it was observed that saturation levels of α-particle irradiated polycrystalline oxides obtained in the experiment were compatible with this order. There is no relation with polycrystalline oxides exposed to the α-decay event because experiments for the same material of different experimental groups were not even consistent with each other. One of the main reasons for this is thermal spikes caused by recoil atoms during the α-decay. The rapid increase in temperature can cause the annihilation of some OT cation defects. The sensitivity

to conditions of material preparation and the possibility of annealing of OT defects are other possible reasons for different findings.

The only material that can be directly compared with experimental data is the single-crystal $UO_2$ structure among the single-crystal structures investigated in this study. The experimental and simulation saturation levels of single-crystal $UO_2$ structures are 0.84% and ~1.03%, respectively. The difference between them is due to the "gentle" placement of defects in the simulation box. This means that at the beginning of each simulation cations were manually placed in interatomic unoccupied positions, and the excess energy is removed by the energy minimization method or by velocity rescaling method [5]. However, in the case of α-radiation, the energy of α-particles is transferred to the crystal as kinetic energy so that a rapid rise in temperature occurs, resulting in the annihilation of some OT defects. A simple computer experiment was designed and simulations without energy minimization were initiated. This method yielded a saturation level of ~0.74%, which is closer to experimental data so that the effect of the excess energy in the irradiated crystal was roughly observed. When the relationship between OT type defects and lattice parameter change was examined in this computer experiment without energy minimization, the existence of a two-part linear equation could still be observed in single-crystal $UO_2$ (see Fig. S14 in the supplementary material). Slopes of the first and the second parts had slightly altered compared to the energy minimized simulation. As screenshots and data were investigated, it was seen that the decrease in the number of OT-type uranium and the increase in the number of oxygen defects had an effect on this change. The data were clustered and scattered throughout the linear equation. The gap between the two linear equations increased. Radial distribution functions have liquid-like appearances at both low and high doses (see Fig. S15, S16 in the supplementary material). If this simple computer experiment can be performed in a more controlled manner, for example, if the amount of extra energy to be transferred to the simulation box can be precisely adjusted, the effect of α-particle radiation can be more accurately modeled. This part requires further research.

$\Delta v_F$ volume increments per OT cation defects were calculated from slopes of lattice dilatation-OT cation defect number two-part linear equation. Unlike the original use of this equation, it was only applied to OT type defects, not to all defects. The OT thorium defect of $ThO_2$ has the largest and the OT plutonium defect of $PuO_2$ has the smallest volume, which is calculated from the first part of the linear equation. Surprisingly, as a result of the calculations from the second part of the linear equations, OT thorium was the defect with almost the smallest volume.

The effect of atom configurations on RDF data was examined for low and high doses of α-particle irradiation. These two samples also represent the primary and secondary parts of the lattice dilatation-OT cation defect number linear equation. Pre- and post-peaks that do not exist in a perfect crystal structure were observed in all oxides in this study. Probably the main reason for this is that they have similar unit cell sizes and the same structures. Since the mass of the Ce atom is quite different from the others (Cm, Am, U, Np, Pu, Th), the similarity between the cation masses cannot be mentioned. The presence of prominent pre-peaks in cation-cation RDF data of all α-particle irradiated oxides indicate the presence of OT cation defects. Positions of pre-peaks are close to each other for most oxides. OT (interstitial) cation defects are located at octahedral sites surrounded by six other cations and octahedrons have a little asymmetrical structure. This asymmetry is likely to increase at high radiation doses. The shape and the distribution of the asymmetric octahedron structure had not been studied in this study and require further investigation because these structures have a direct impact on the lattice dilation. One reason for this asymmetry may be the exposure of MD simulation boxes to the constant stress method. Both MD simulation box shape and size can be changed by this technique. In future studies, structures of octahedron cages should also be examined by applying constant pressure methods to the MD boxes.

The $ThO_2$ pre-peak is located on the x-axis of RDF at a different position from pre-peaks of other oxides for both low and high doses of α-particles. There are also post-peaks in RDF data, and these peaks emerge as a result of OT type defects pushing other cations (DT type defects) into the <110> channel. Apparent boundaries of RDF pre-peaks provide more accurate OT defect number calculations, but as uncertainties in locations of boundaries increase, the number of OT defect data become more dispersed. The uncertainty in the location of the RDF peak boundary increases with increasing radiation doses.

## 5. Conclusion

Imperfections and temperature have a direct effect on the change of lattice constants of molecules. In this study, the effects of OT-type defects on some actinide and lanthanide dioxides with the fluorite-type structure were discussed and their importance was revealed. The swelling of the lattice is determined as a result of the delicate balance between OT defect number and temperature (e.g., thermal spikes). In order to observe the extent of effects of these defects more clearly, computer tests involving other defects (e.g., clusters, radiogenic Pb, He, etc.) should be performed. Studies should also be made on polycrystalline actinide and lanthanide dioxides and effects of OT defects on these structures should be investigated. The relationship between OT defects with other physical quantities, like thermal conductivity, diffusion, ionic conductivity, heat capacity, etc., should also be studied so that some events that have not been fully explained so far can be solved, or a new perspective can be provided for the already described events. Research should be extended by adding the effect of temperature to all these studies. The study should be further expanded and a more general theory should be put forward by studying all fluorite-structured oxides or similar molecules.

**Acknowledgements**

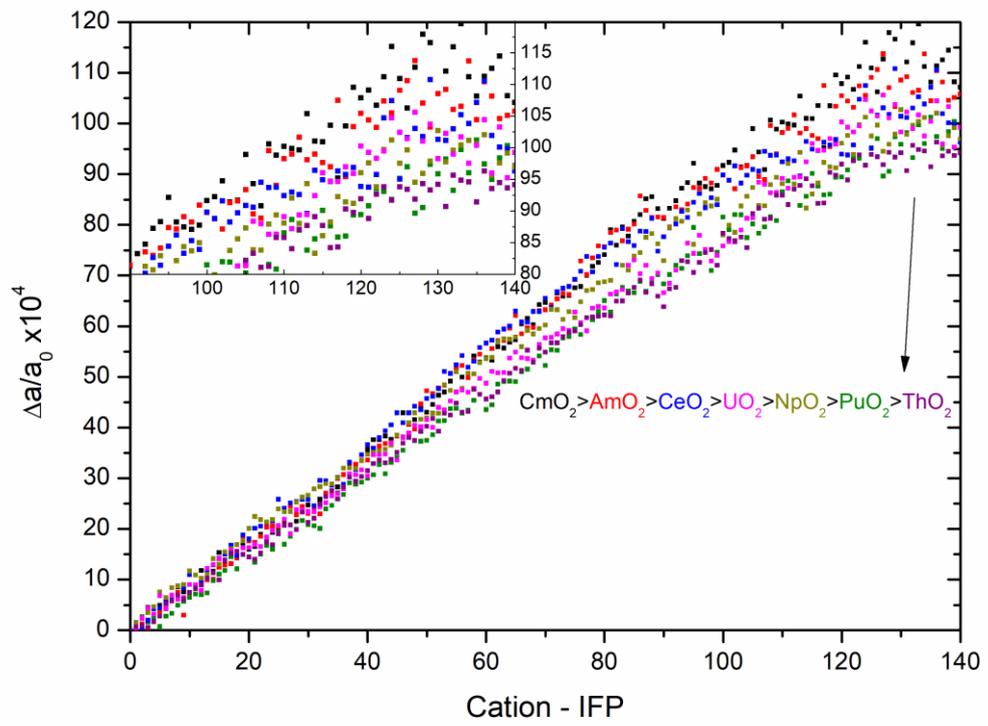

**Fig. 1.** The variation of changes in lattice parameters with the initial number of Frenkel pairs.

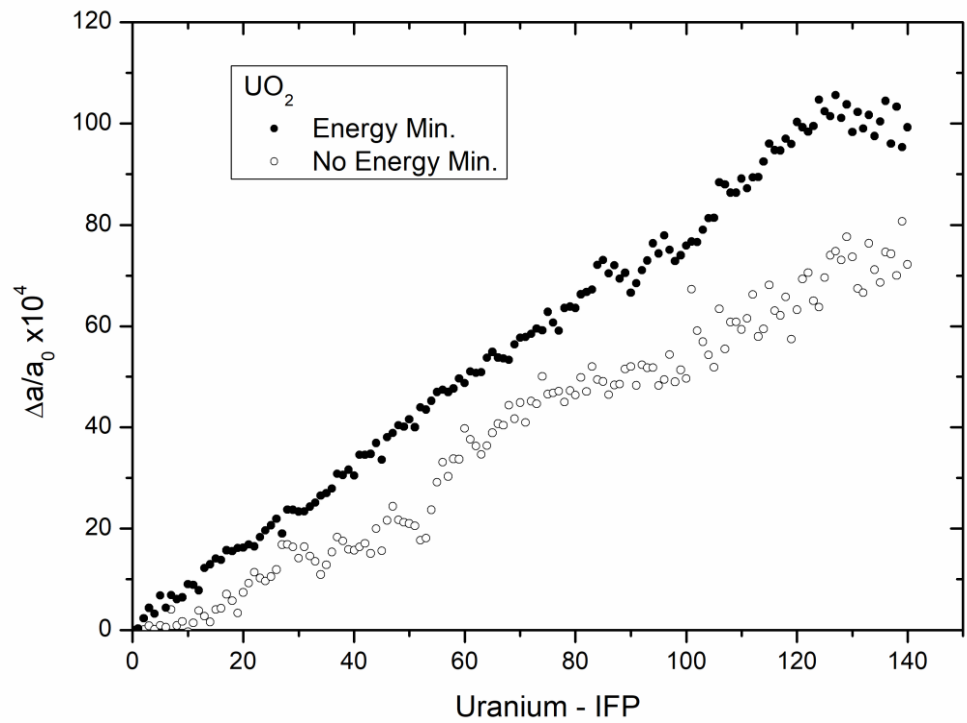

**Fig. 2.** The lattice parameter change of $UO_2$ versus the initial number of uranium Frenkel pairs with and without energy minimization methods.

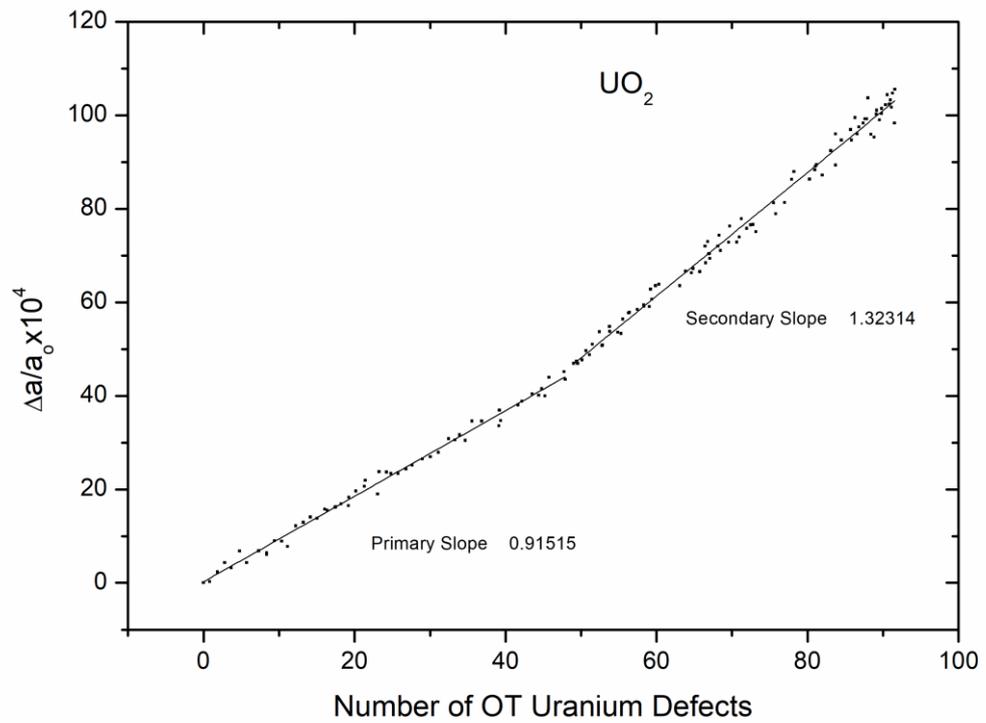

**Fig. 3.** The lattice parameter change in $UO_2$ with the number of OT uranium defects.

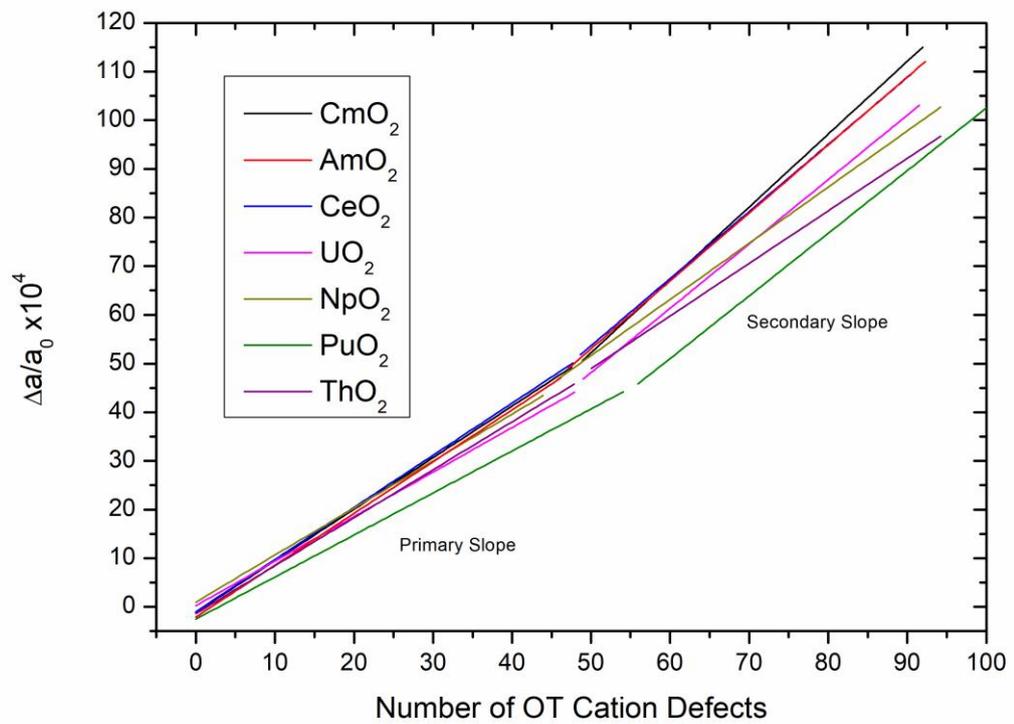

**Fig. 4.** Lattice parameter changes with OT cation defects.

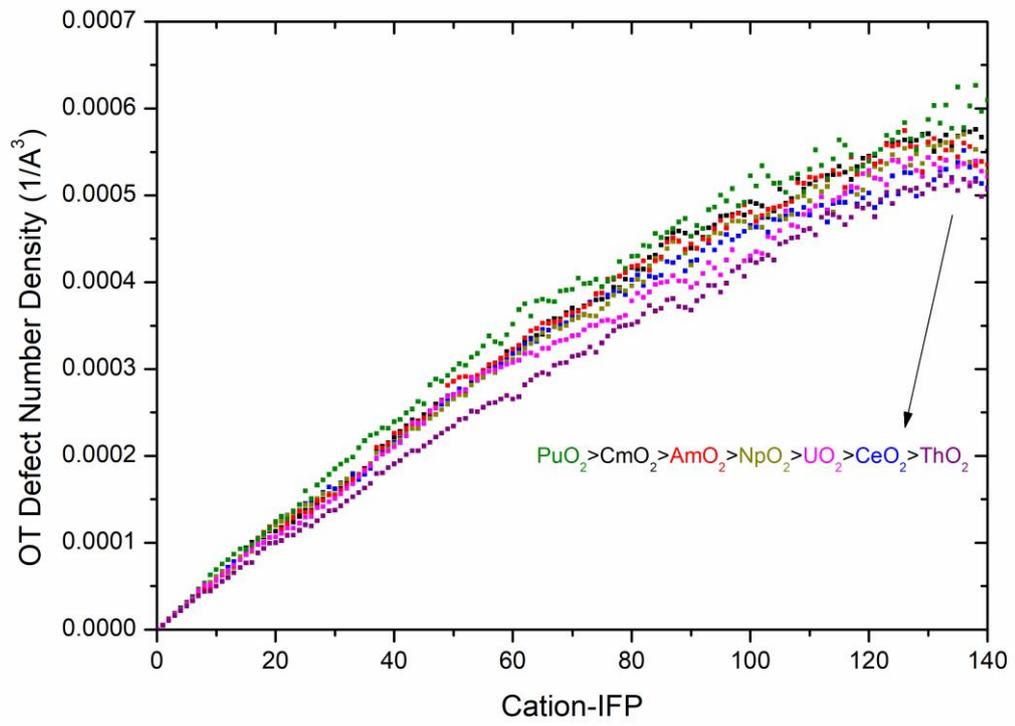

**Fig. 5.** OT defect number density data versus IFP cation data.

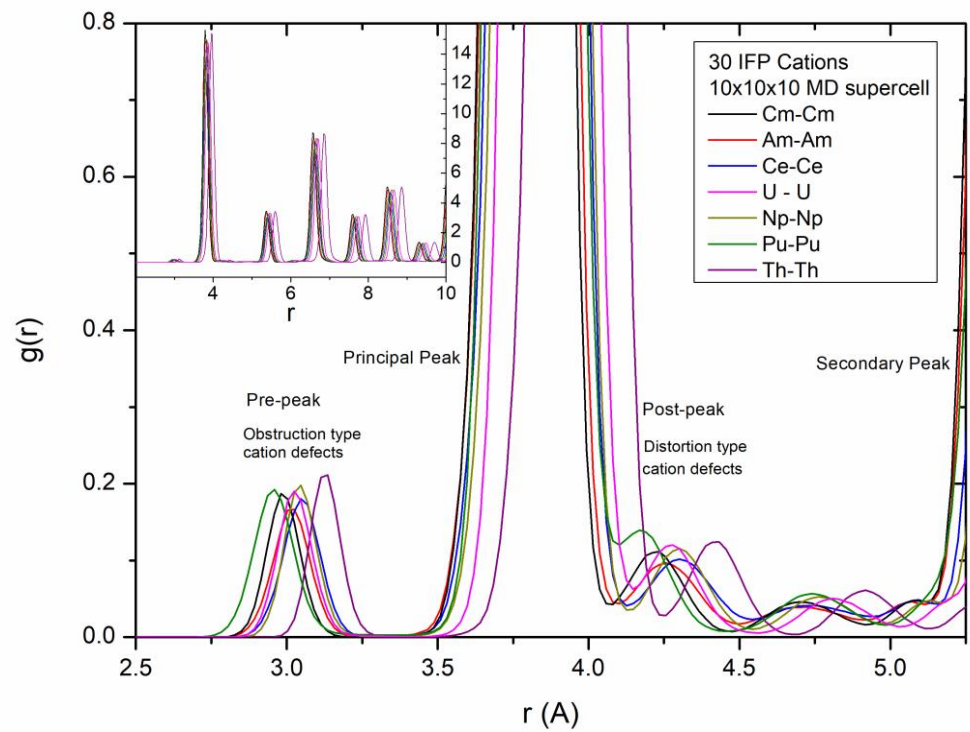

**Fig. 6.** Cation-cation radial distribution functions of simulation boxes with 30 IFP cations.

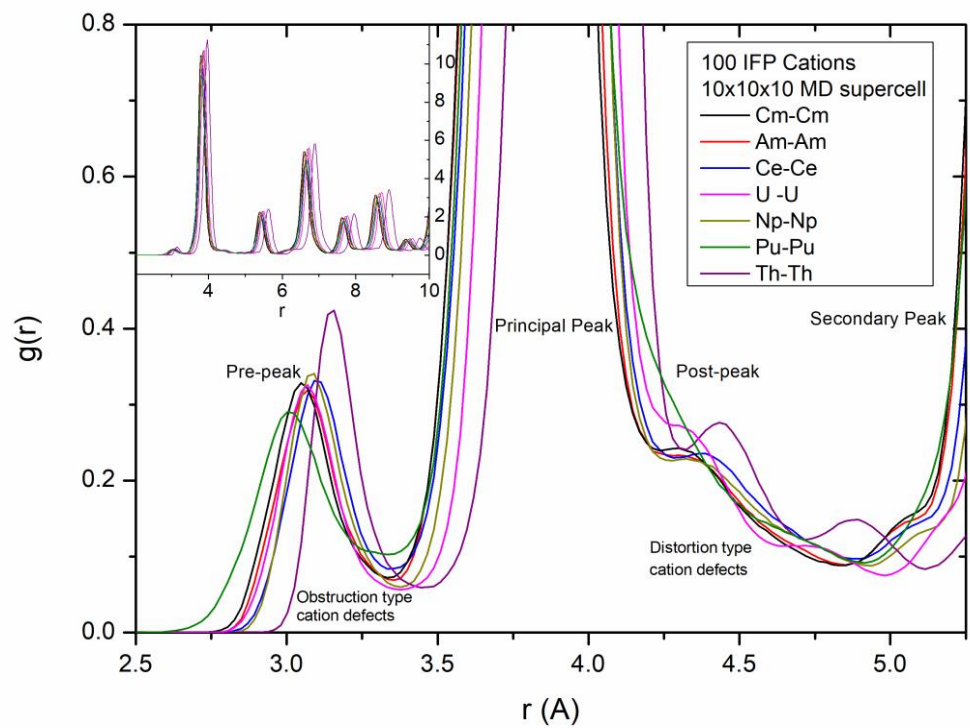

**Fig. 7.** Cation-cation radial distribution functions of simulation boxes with 100 IFP cations.

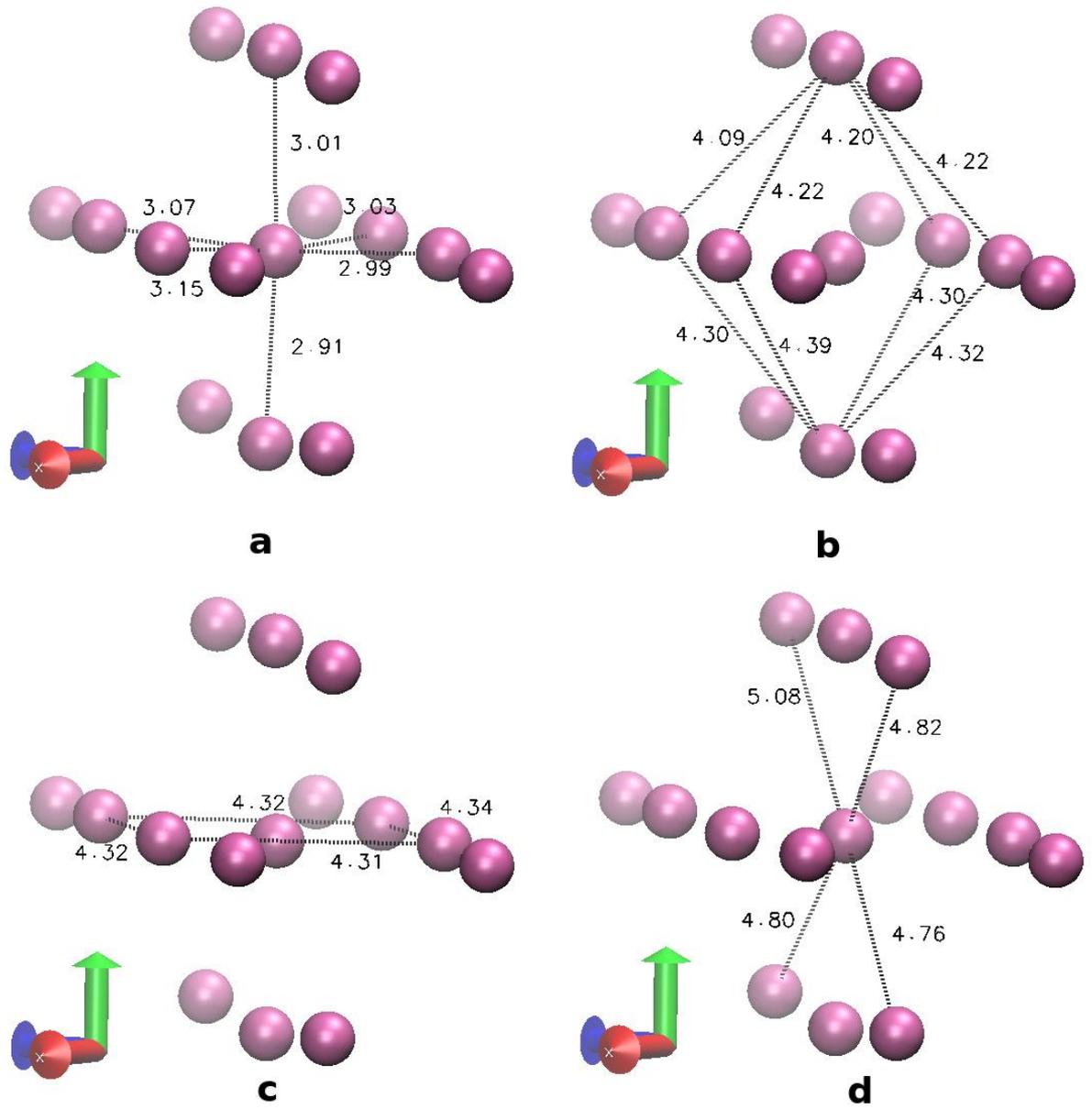

**Fig. 8.** OT uranium defect is visualized from the <110> direction in 30 IFP sample. The figure is slightly tilted about the y-axis and oxygen atoms were removed from the figure for reasons of better viewing and clarity.

**Table 1.** The saturation point values of change in lattice parameters were listed. The second column gave the average value of single-crystal structures from MD. In the third and the fourth column, experimental data on polycrystalline and single-crystal materials were given, respectively.

| Molecule | $(\Delta a/a_0)_{sat}$ (MD, Single-Crys) | $(\Delta a/a_0)_{sat}$ (Exp, Poly-Crys, α-Irr) [3] | $(\Delta a/a_0)_{sat}$ (Exp, Single-Crys, α-Irr) [1] |
|---|---|---|---|
| $CmO_2$ | ~1.14% | | |
| $AmO_2$ | ~1.09% | | |
| $CeO_2$ | ~1.05% | 0.73% | |
| $UO_2$ | ~1.03% | 0.58% | 0.84% |
| $NpO_2$ | ~1.00% | | |
| $PuO_2$ | ~0.98% | 0.46% | |
| $ThO_2$ | ~0.94% | | |

**Table 2.** Comparison of single-crystal oxides MD simulation results with self-radiation damage experiments due to α-decay (α-particle and α-recoil atom bombardment) on polycrystalline oxides in the literature.

| Molecule | $(\Delta a/a_0)_{sat}$ (MD, Single-Crys) | $(\Delta a/a_0)_{sat}$ (Exp, Poly-Crys, Self-Rad) |
|---|---|---|
| **CmO$_2$** | ~1.14% | 0.233%[13] |
| | | 0.3 % [8] |
| **AmO$_2$** | ~1.09% | 0.282%[19] |
| | | 0.338%[9] |
| | | 0.35% [20] |
| | | 0.239% [10] |
| **CeO$_2$** | ~1.05% | |
| **UO$_2$** | ~1.03% | |
| **NpO$_2$** | ~1.00% | |
| **PuO$_2$** | ~0.98% | 0.283% [13] |
| | | 0.338% [9] |
| | | 0.39% [11] |
| | | 0.32% [21] |
| **ThO$_2$** | ~0.94% | |

**Table 3.** Second and third columns were slopes of lattice parameter changes with OT cation defects function. Fourth and fifth columns were volume increments per OT cation defects. The last column was the difference between the fifth and fourth columns.

| Molecule | Primary Slope($\times 10^4$) | Secondary Slope($\times 10^4$) | Primary $\Delta v_{F1}(\text{Å}^3)$ | Secondary $\Delta v_{F2}(\text{Å}^3)$ | Difference $\Delta v_{F2} - \Delta v_{F1}$ |
|---|---|---|---|---|---|
| $CmO_2$ | 1.0598 | 1.49706 | 49.042 | 69.276 | 20.234 |
| $AmO_2$ | 1.06424 | 1.39613 | 49.601 | 65.077 | 15.476 |
| $CeO_2$ | 1.07181 | 1.37902 | 50.997 | 65.615 | 14.618 |
| $UO_2$ | 0.91515 | 1.32314 | 44.909 | 64.931 | 20.022 |
| $NpO_2$ | 0.96595 | 1.15461 | 46.447 | 55.518 | 9.071 |
| $PuO_2$ | 0.86383 | 1.28739 | 40.716 | 60.680 | 19.964 |
| $ThO_2$ | 0.98595 | 1.07999 | 51.806 | 56.747 | 4.941 |

**Table 4.** Tabulated data were obtained from lattice parameter changes versus OT cation defect number data for each molecule.

| Molecule | Break Point (x,y) | Maximum Point (x,y) |
|---|---|---|
| $CmO_2$ | (48, 0.50%) | (91, 1.20%) |
| $AmO_2$ | (42, 0.42%) | (90, 1.13%) |
| $CeO_2$ | (45, 0.48%) | (90, 1.10%) |
| $UO_2$ | (48, 0.45%) | (91, 1.05%) |
| $NpO_2$ | (44, 0.43%) | (94, 1.02%) |
| $PuO_2$ | (54, 0.46%) | (101, 1.01%) |
| $ThO_2$ | (47, 0.47%) | (93, 0.96%) |

# Supplementary Material

# Actinide and Lanthanide Dioxide Lattice Dilatation Mechanisms with Defect Ingrowth

Seçkin D. Günay

*Department of Physics, Faculty of Science, Yıldız Technical University, Esenler, 34210, İstanbul, Turkey*

Corresponding author. Tel.: +90 212 3834289 E-mail address: sgunay@yildiz.edu.tr

**Orders of data obtained from the molecular dynamics simulation calculations of single-crystal (MD) and experimental results of polycrystalline (EXP) isostructural oxides from large to small values are as follows:**

- MD: Saturation levels of α-particle irradiated single-crystal structures (see Table 1)

  $CmO_2$, $AmO_2$, $CeO_2$, $UO_2$, $NpO_2$, $PuO_2$, $ThO_2$

- EXP: Saturation levels of α-particle irradiated structures [3] (see Table 1)

  $CeO_2$, $UO_2$, $PuO_2$

  Although Weber [3] did not perform an experiment on the polycrystalline $ThO_2$ molecule, he predicted that the molecule should be first in the above list, which is inconsistent with our data.

- MD: The primary slope of lattice dilatation versus the number of OT cation defects (see Table 3),

  $CeO_2$, $AmO_2$, $CmO_2$, $ThO_2$, $NpO_2$, $UO_2$, $PuO_2$

- MD: The secondary slope of lattice dilatation versus the number of OT cation defects (see Table 3),

  $CmO_2$, $AmO_2$, $CeO_2$, $UO_2$, $PuO_2$, $NpO_2$, $ThO_2$

- MD: Volumetric increases per OT cation defects $\Delta v_F$ calculated from the primary slope of lattice dilatation versus the number of OT cation defects (see Table 3),

  $ThO_2$, $CeO_2$, $AmO_2$, $CmO_2$, $NpO_2$, $UO_2$, $PuO_2$

- MD: Volumetric increases per OT cation defects $\Delta v_F$ calculated from the secondary slope of lattice dilatation versus the number of OT cation defects (see Table 1),

  $CmO_2$, $CeO_2$, $AmO_2$, $UO_2$, $PuO_2$, $ThO_2$, $NpO_2$

- MD: Difference between second and first volumetric increases per OT cation defect (see Table 3),

  $CmO_2$, $UO_2$, $PuO_2$, $AmO_2$, $CeO_2$, $NpO_2$, $ThO_2$

- EXP: Volumetric increases per Frenkel defect pair [3],

  $CeO_2$, $UO_2$, $PuO_2$

  This ranking in Weber's work is given for information purposes only, and it would not be correct to compare sequences in experimental and simulation data due to the difference between definitions of volumetric increase. See section 3.1.3 for more information.

- MD: The number of OT cation defect-bearing capacities (OT defect number density) at the saturation level (see Fig. 5)

PuO₂, CmO₂, AmO₂, NpO₂, UO₂, CeO₂, ThO₂

- EXP: The number of defects per unit cell [3].

PuO₂, UO₂, CeO₂

**Other general information on actinides and lanthanides from large to small values are as follows:**

- EXP: Ionic radius $M^{+4}$

Th, U, Ce=Np, Pu, Am = Cm

- EXP: Lattice parameters of perfect isostructural oxides

ThO₂, UO₂, NpO₂, CeO₂, PuO₂, AmO₂, CmO₂

There is no relationship between the order of the last two experimental values and the order of the values obtained by MD simulation method.

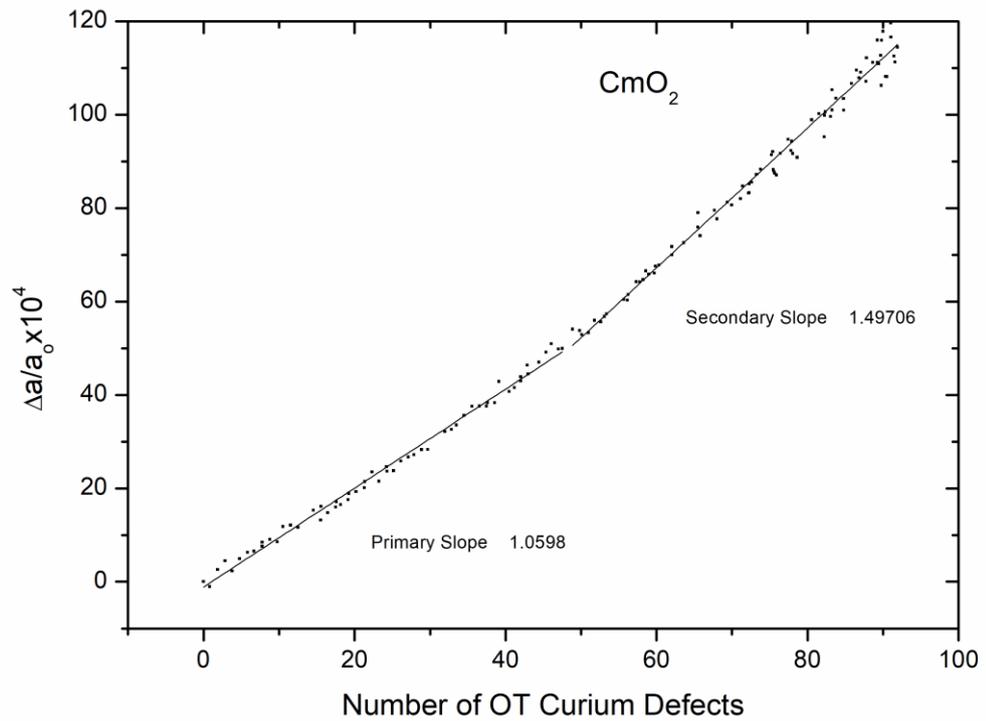

**Fig. S1.** The lattice parameter change in CmO₂ with the number of OT curium defects.

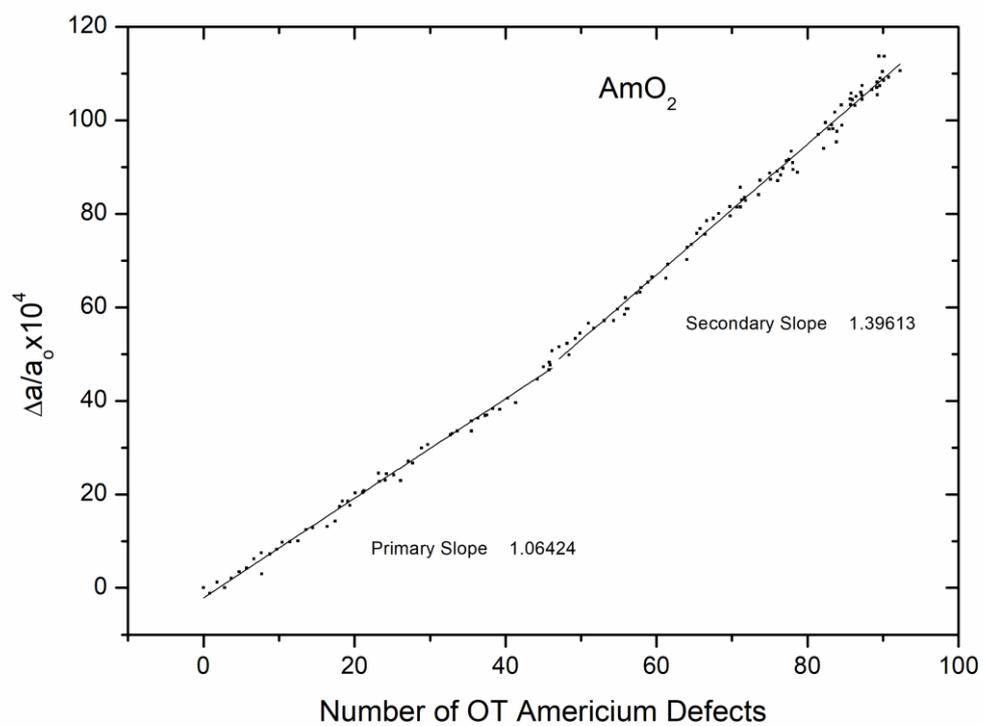

**Fig. S2.** The lattice parameter change in AmO$_2$ with the number of OT americium defects.

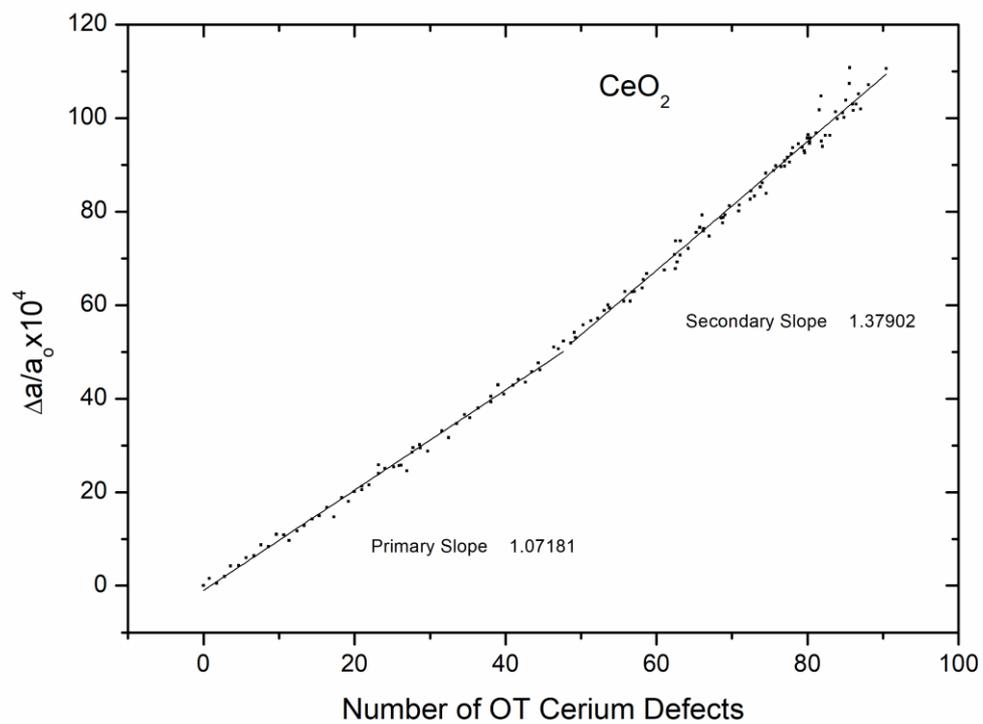

**Fig. S3.** The lattice parameter change in CeO$_2$ with the number of OT cerium defects.

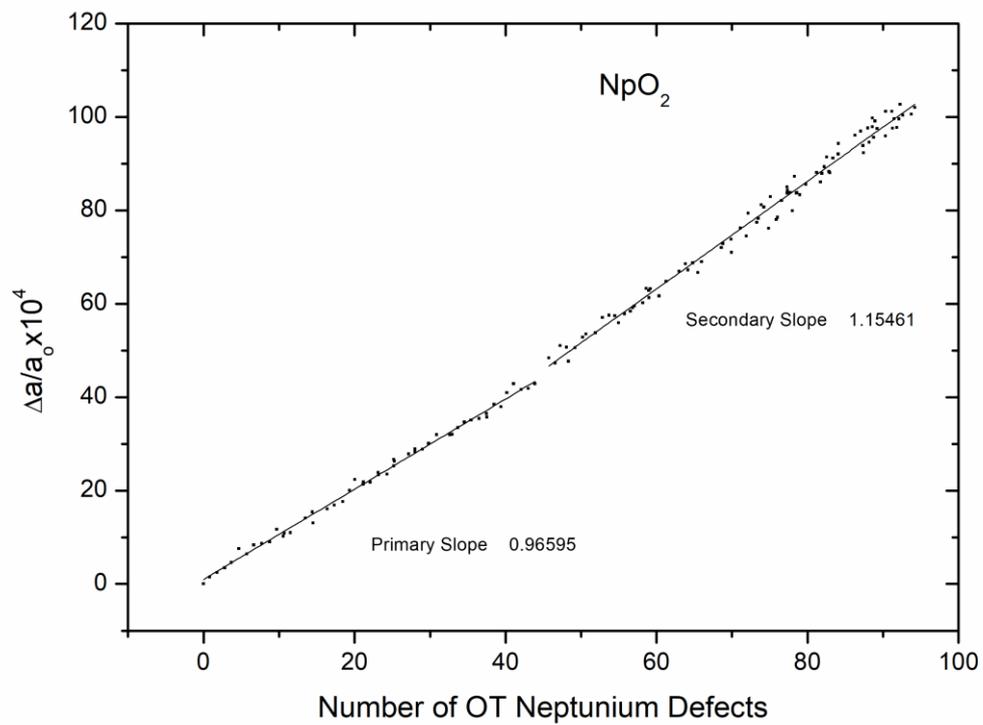

**Fig. S4.** The lattice parameter change in NpO$_2$ with the number of OT neptunium defects.

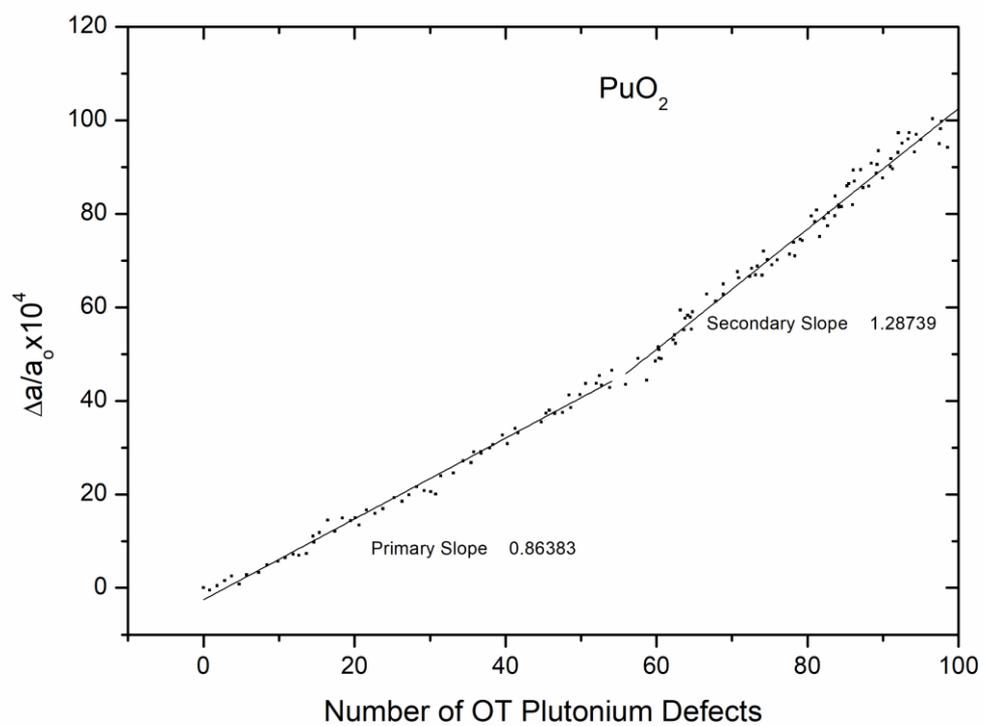

**Fig. S5.** The lattice parameter change in PuO$_2$ with the number of OT plutonium defects.

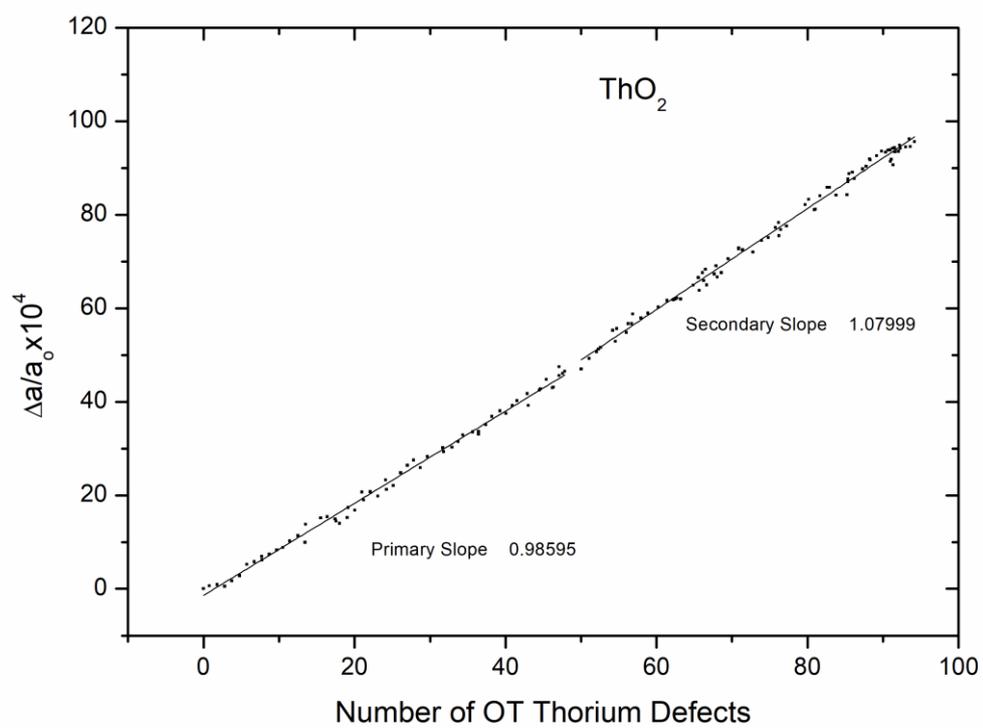

**Fig. S6.** The lattice parameter change in ThO$_2$ with the number of OT thorium defects.

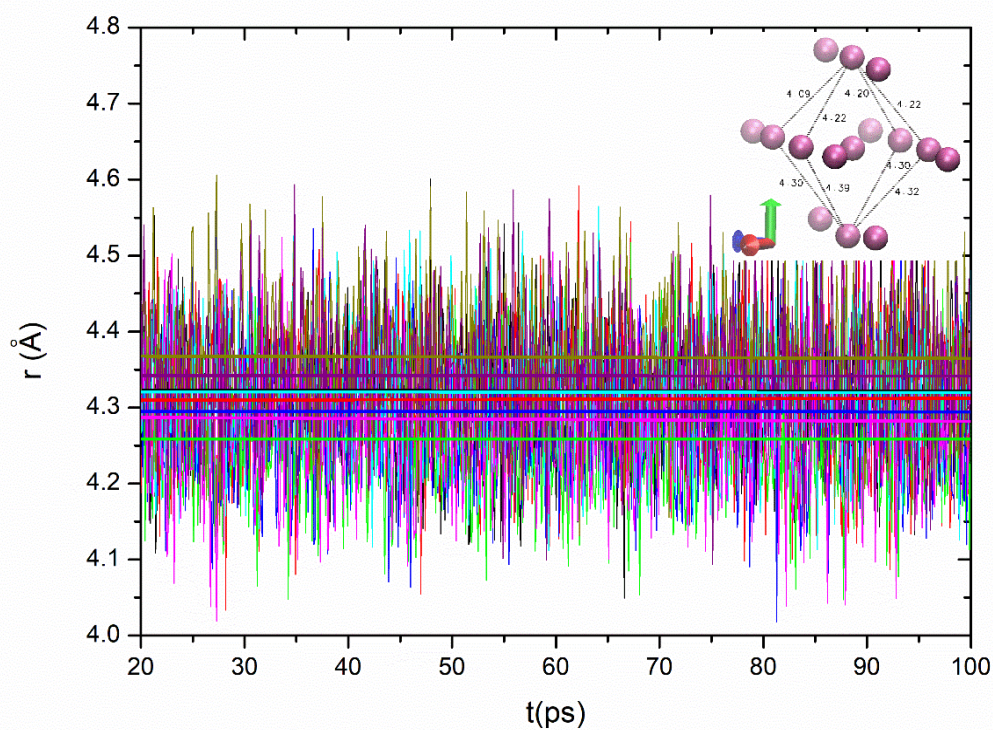

**Fig. S7.** Uranium - uranium distances of an octahedron surrounding OT defect versus time. Linear equations were fitted to each of distance data over time and equations were represented by colored lines. OT uranium defect is in the middle of the six uranium atoms.

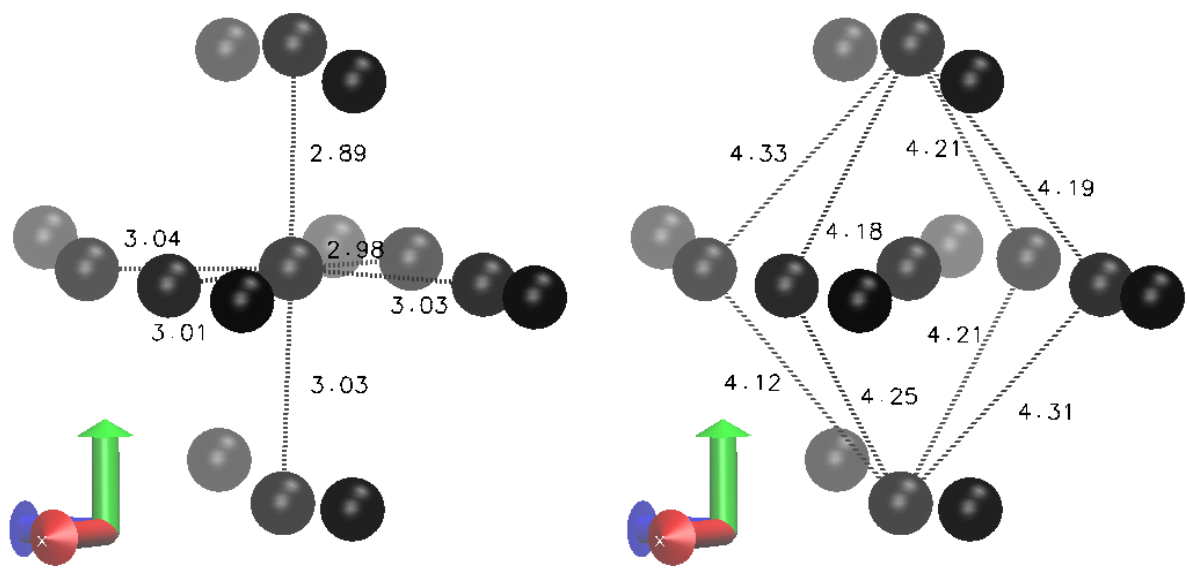

**Fig. S8.** OT curium defect is visualized from the <110> direction in 30 IFP sample. The figure is slightly tilted about the y-axis.

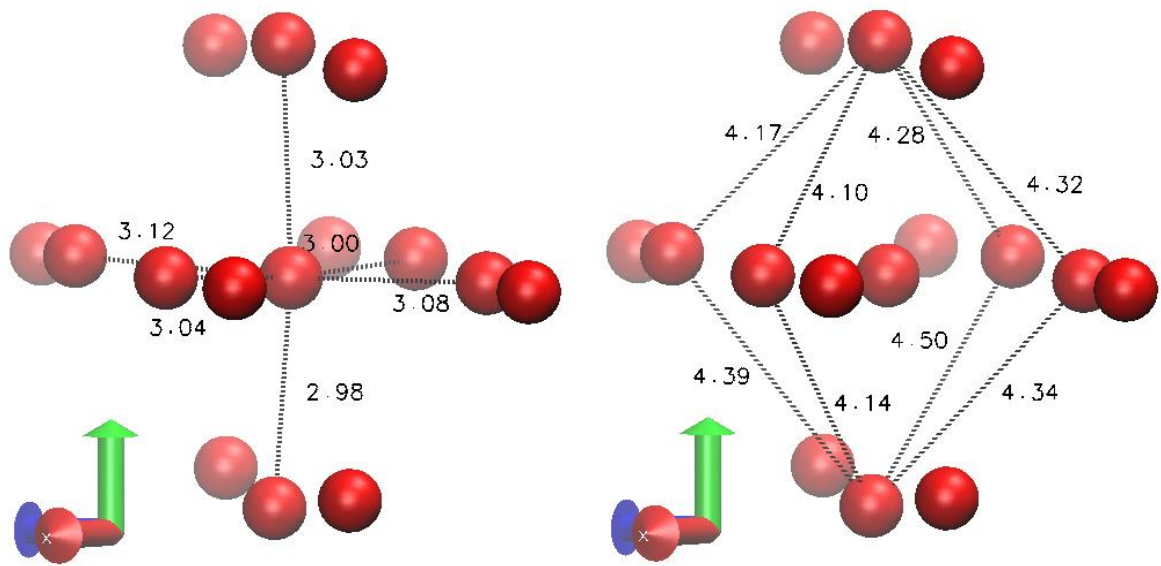

**Fig. S9.** OT americium defect is visualized from the <110> direction in 30 IFP sample. The figure is slightly tilted about the y-axis.

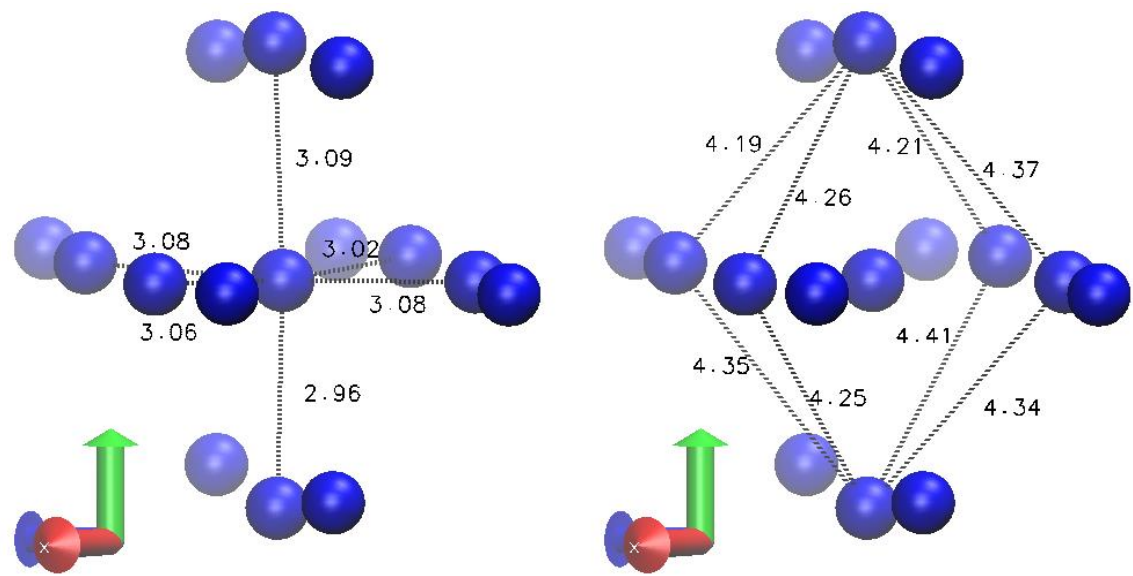

**Fig. S10.** OT cerium defect is visualized from the <110> direction in 30 IFP sample. The figure is slightly tilted about the y-axis.

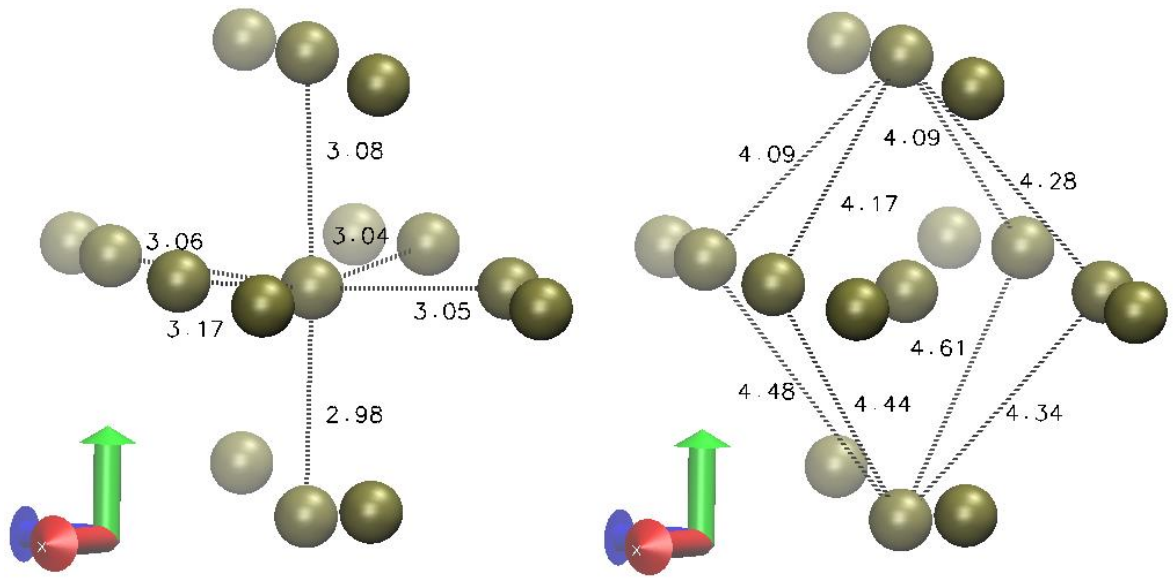

**Fig. S11.** OT neptunium defect is visualized from the <110> direction in 30 IFP sample. The figure is slightly tilted about the y-axis.

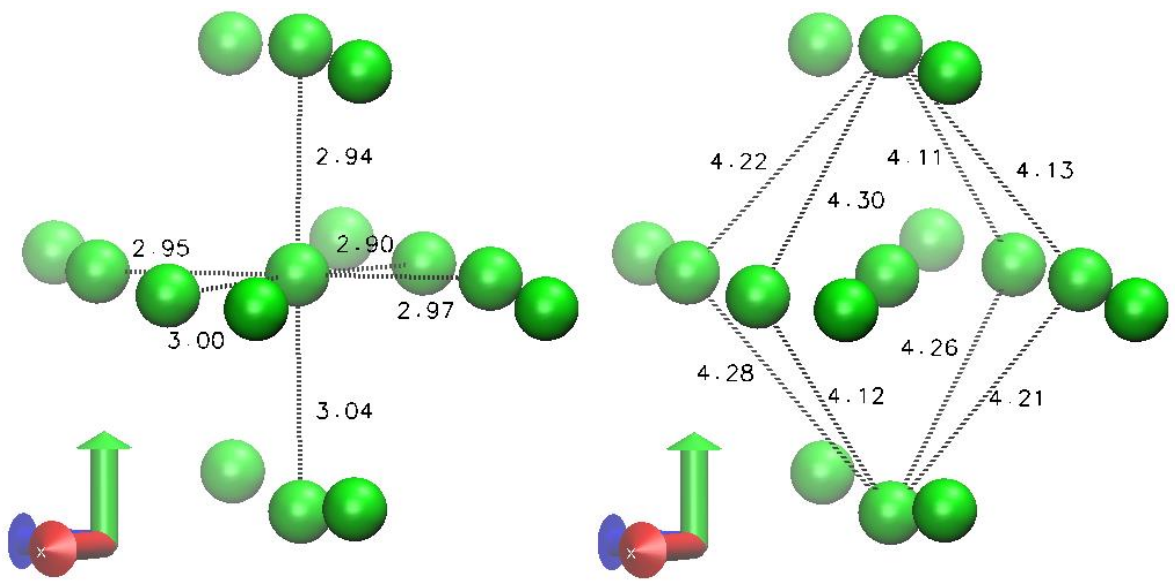

**Fig. S12.** OT plutonium defect is visualized from the <110> direction in 30 IFP sample. The figure is slightly tilted about the y-axis.

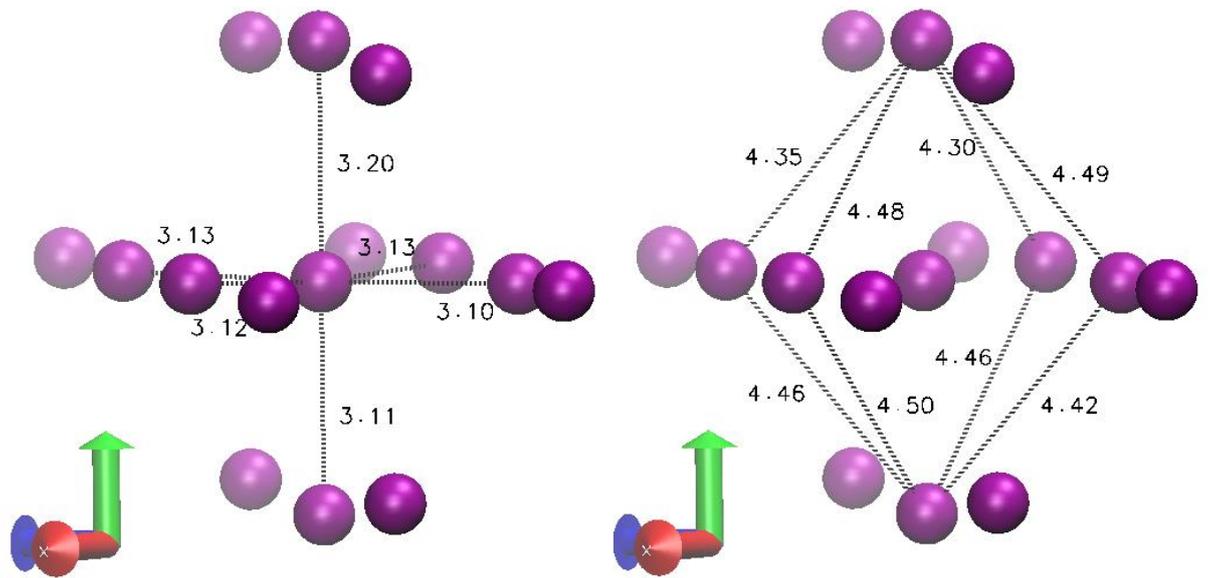

**Fig. S13.** OT thorium defect is visualized from the <110> direction in 30 IFP sample. The figure is slightly tilted about the y-axis.

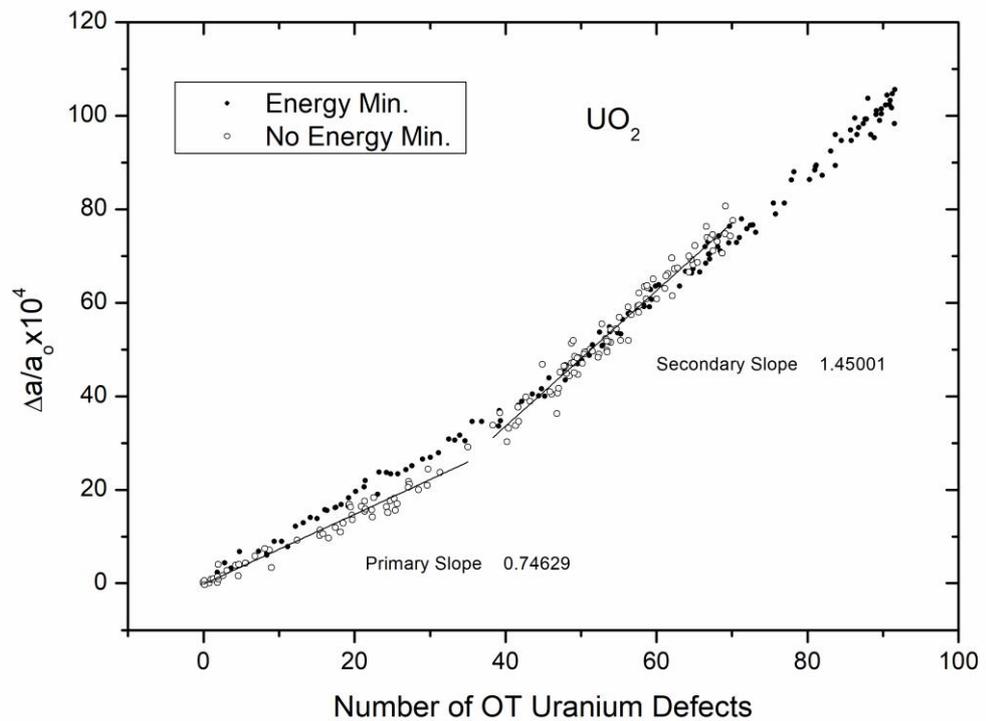

**Fig. S14.** The lattice parameter change of $UO_2$ versus the number of OT uranium defects with and without energy minimization methods.

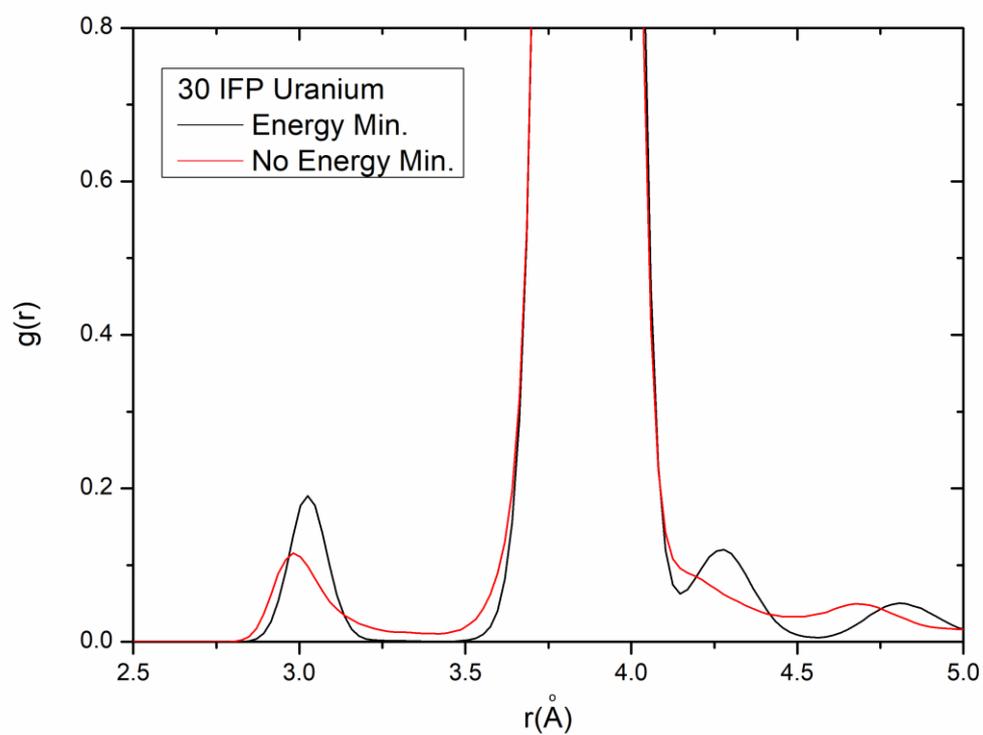

**Fig. S15.** Radial distribution functions of UO$_2$ simulation boxes with 30 IFP uranium defects.

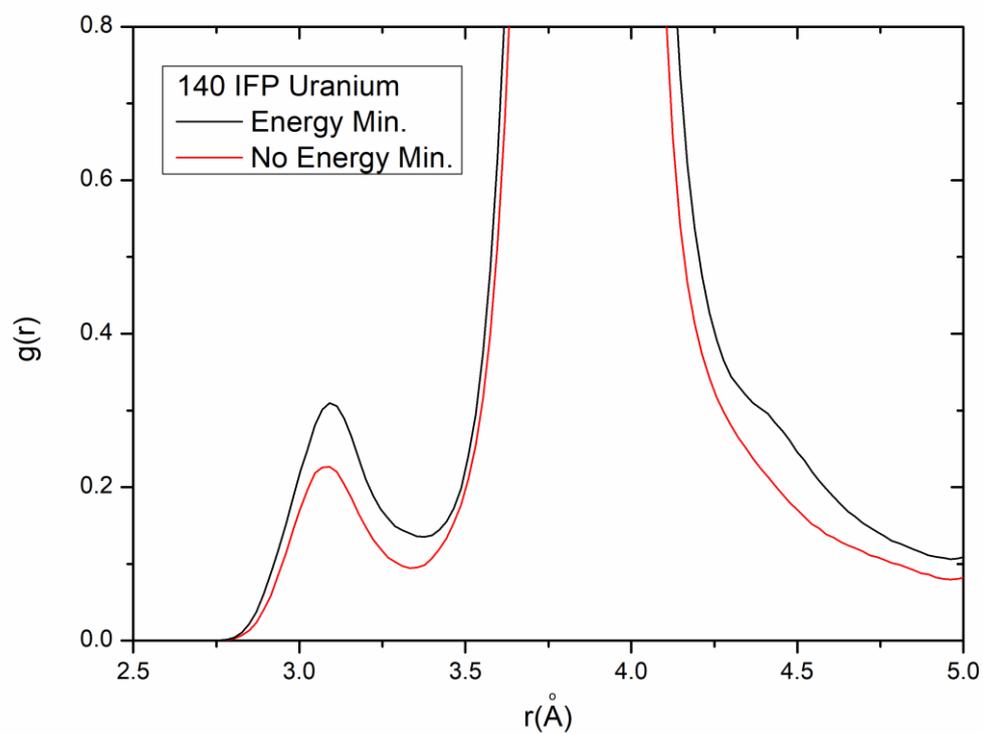

**Fig. S16.** Radial distribution functions of UO$_2$ simulation boxes with 140 IFP uranium defects.